\documentclass[pre,showpacs,twocolumn,superscriptaddress,showkeys]{revtex4}
\usepackage{amssymb,graphicx}

\begin{document}
\title{Dissolution of traffic jam via additional local interactions}

\author{Hyun Keun Lee}
\affiliation{BK21 Physics Research Division and Department of Physics, Sungkyunkwan University, Suwon 440-746, Korea}
\affiliation{Department of Physics, University of Seoul, Seoul 130-743, Korea}
\author{Beom Jun Kim}
\email[Corresponding author, Tel: +82-31-299-4541, Fax: +82-31-290-7055, E-mail: ]{beomjun@skku.edu}
\affiliation{BK21 Physics Research Division and Department of Physics, Sungkyunkwan University, Suwon 440-746, Korea}
\affiliation{Asia Pacific Center for Theoretical Physics, Pohang 790-784, Korea}

\begin{abstract}
We use a cellular automata approach to numerically investigate traffic flow
patterns on a single lane. The free-flow phase (F), the synchronized phase (S),
and the jam phase (J) are observed and the transitions among them are studied
as the vehicular density $\rho$ is slowly varied.  If $\rho$ is decreased from
well inside the J phase, the flux $\Phi$ follows the lower branch of the
hysteresis loop, implying that the adiabatic decrease of $\rho$  is not an
efficient way to put the system back into S or F phases.  We propose a simple
way to help the system to escape out of J phase, which is based on the local
information of the velocities of downstream vehicles.
\end{abstract}
\keywords{Traffic phases; Cellular automata; Traffic jam; Local interaction; Hysteresis; Phase transition}
\pacs{05.45.-a,05.60.-k,05.65.+b}
\maketitle

Since 1990s, traffic phenomena have drawn much attention amongst physicists~\cite{review}.  Pure theoretical interest of the study of the traffic
problem originates from its variety of fascinating dynamic emergent
behaviors~\cite{HK-empirical}, like the phase transitions between different
dynamical states, the existence of hysteresis behavior, and the historic
dependence of traffic flow, to name a few.  Recently, there has been a
growing consensus on the existences of three different dynamic
phases: when the vehicular density $\rho$ is small enough, the
system first stays in the free-flow (F) phase in which vehicles can have the
maximum speed allowed, and thus the flux $\Phi$ is linearly proportional to
$\rho$.  As $\rho$ is increased, the speeds of vehicles are synchronized (S phase)
to each other, without having stop-and-go motion, and then eventually the jam (J)
phase arises beyond some value of $\rho$.

\begin{figure}
\includegraphics[width=0.4\textwidth]{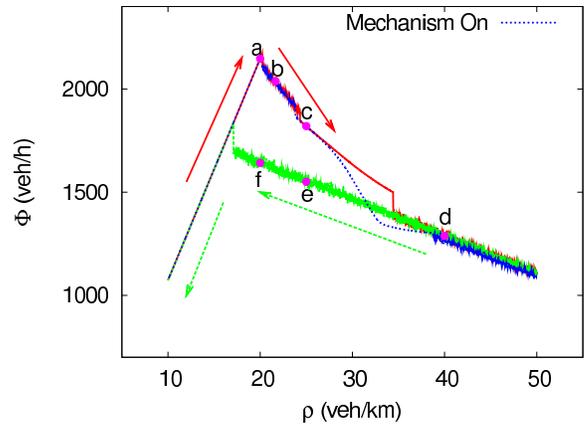}
\caption{(Color online) Fundamental diagram: the flux $\Phi$ versus the vehicular
density $\rho$ shows a hysteresis behavior.  As $\rho$ is increased from the
free-flow (F) phase  ({\bf a}), $\Phi$ follows the upper branch of the curve, passing
the synchronized (S) phase ({\bf b} and {\bf c}), and the jam (J) phase ({\bf
d}).  When started from the J phase ({\bf d}), $\Phi$ follows the lower
branch of the curve, passing the points {\bf e} and {\bf f} (both are in J
phase, see Fig.~\ref{fig:xt}). The dotted blue curve in the middle is obtained
when our method of dissolution of the jam by using an additional interaction is turned on.
}
\label{fig:rhoPhi}
\end{figure}

The cellular automata (CA) model proposed by Kerner et al. (KKW)~\cite{KKW} 
to explain the three traffic phases, has been recently applied for 
multilane roads and compared with single vehicle data measured
on American highway~\cite{kk2009}. 
Knospe {\it et. al.}~\cite{knospe} extended Nagel-Schreckenberg
model~\cite{NaSch} to include the anticipation effect and the reduced
acceleration capabilities, and a further extension was made by Jiang and
Wu~\cite{jiang} with the difference of sensitivities between stopped and moving
cars taken into account. 
Davis~\cite{davis} used the modified optimal velocity
model and also produced the synchronized flow phase, but with unrealistic speed
of upstream front.  In Ref.~\cite{hklee}, realistic assumptions of the limited 
deceleration capability and human overreaction has been used as important
ingredients of the CA model.
Similar ideas of the limited decelerating capability and  cautious driving when congested
have been adopted in Ref.~\cite{yang}, in which a further improvement was made
by letting the maximum speed be dependent on headway distance to incorporate
the idea of the comfortable driving.  A CA model with the similar deceleration
limitation but without the slow-to-start effect has been proposed, which again
reproduced the three traffic phases and discontinuous transitions between
them~\cite{jin}, which in turn suggests that the slow-to-start effect is not
a cause but a result of the transition to the wide-moving jam.  
In Ref.~\cite{xgli}, an attempt to interpret the
traffic flow in the viewpoint of the complex network of traffic states was
made.  
Very recently, the emergence of a jam without bottlenecks has
been observed experimentally~\cite{sugiyama}, which clearly indicates that the
traffic jam is a collective phenomenon resulting from interactions between
vehicles. 
In the present paper, we follow the same line of reasoning and propose that
one can use an additional local interaction to dissolve the traffic jam.  

\begin{figure*}
\includegraphics[width=0.9\textwidth]{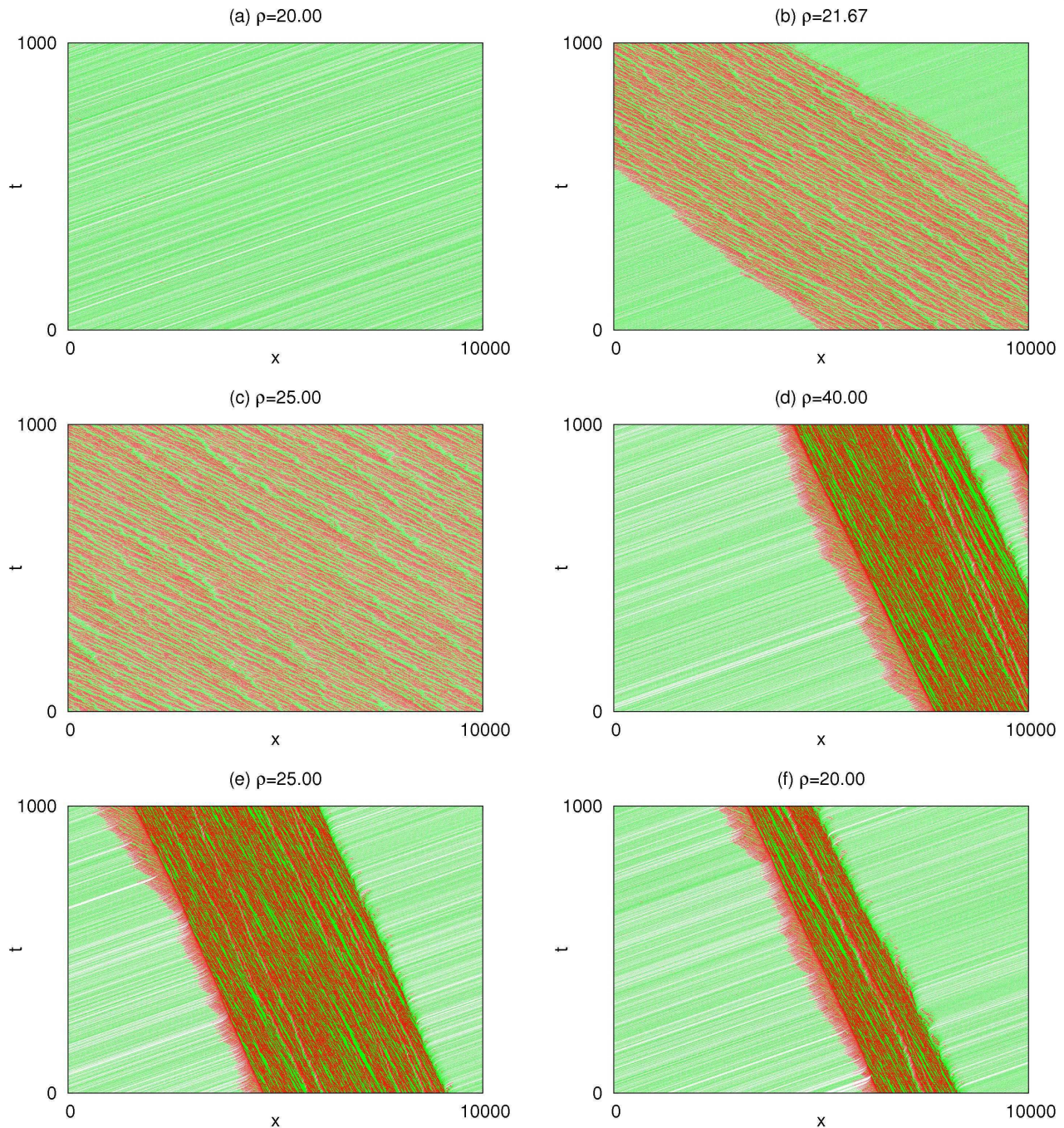}
\caption{(Color online) Spatio-temporal ($x$,$t$) patterns of traffic flows at points {\bf a} - {\bf f} in Fig.~\ref{fig:rhoPhi}.
Dots with bright and the dark colors (green and red online, respectively) represent vehicles at
optimistic and defensive states. As the vehicular density $\rho$
is increased (see Fig.~\ref{fig:rhoPhi}), transitions occur
from (a) the F phase to (c) the homogeneous S phase and then to (d) the J phase. Between
the F and the homogeneous S phase, the inhomogeneous S phase in (b) is observed.
As $\rho$ is slowly decreased, the system does not escape out of the J phase
[compare (e) and (c), and (f) and (a), respectively].
}
\label{fig:xt}
\end{figure*}

The main purpose of the present work is not to propose another traffic model,
but to study how the traffic jam can be dissolved through the use of the local
information of car traffic. In this regard, we use the same CA model 
as in Ref.~\cite{hklee}, which has the mechanical
restriction (implemented as limited acceleration and deceleration capabilities)
and the human overreaction as two important ingredients of the model.
Throughout the present study, discretized one unit of time (distance)
amounts to $\Delta t = $1 s ($\Delta x=$1.5 m), and  $x_n^t (v_n^t)$ denotes the
position (the velocity) of the $n$th vehicle at time $t$. The human
driving behavior is described by the two-state variable $\gamma_n^t$,
which is set to $\gamma_n^t = 0$ for $v_n^t \leq v_{n+1}^t \leq v_{n+2}^t$. 
We also assign $\gamma_n^t = 0$ for $v_{n+2}^t \geq v_{\rm max}-1$ 
with the speed limit $v_{\rm max}$. 
If neither condition above is satisfied, we set $\gamma_n^t = 1$ instead.
In words, if the speed is successively nondecreasing in the downstream direction
the driver expects that the traffic condition is improving. She can also be optimistic when the second leading vehicle is
fast enough. The variable $\gamma$ is not quenched but can change in time,
and describes the driving behavior of each driver.
It is to be noted that both $\gamma = 0$ and $\gamma = 1$ represent human
overreaction and only the type of overreaction is different (optimistic
and defensive).

We sketch here only the basic ideas of the model in Ref.~\cite{hklee}:
a driver in the optimistic (defensive) state with $\gamma = 0(1)$ leaves
less (more) distance to the leading vehicle than that required for her safety.
In comparison, an optimistic driver tries to avoid a collision only up to some
time steps.  By introducing optimistic
and defensive behaviors, human factor is modeled as an overreaction to the local
traffic condition.
It is also assumed that vehicles can neither exceed the speed limit $v_{\rm
max} = 20$ (corresponding to 108 km/h), nor drive in the opposite
direction ($v < 0$). As another important ingredient of the model, each vehicle
can decrease their velocity stochastically, to implement the random driving
behavior in reality. All the parameter values are set identical to those in
Ref.~\cite{hklee} to mimic empirical observations. 
In Ref.~\cite{sugiyama}, the critical vehicular density separating the free-flow 
phase the congested flow phase for the experimental setup of circularly moving cars was
found to be about 25 vehicles/km as observed in real highway traffic, and the
speed of the downstream jam front was measured as 20km/h.  
In the present study, for example, the downstream
speed of the wide moving jam is found to be 18.4km/h, similar to the empirical
measurement.

We first report our simulation result~\cite{footnote}
for the fundamental diagram.
We use the single lane of the size
$L=$40 000 (corresponding to 60 km) under the periodic boundary condition.
As an initial condition we uniformly distribute $N= 600$ vehicles
[$\rho = 10$ veh/km] and run the simulation by using the model
in Ref.~\cite{hklee}. In order to neglect transient behaviors, we ignore
the first $T$ time steps and measure the average speed
$v_{\rm avg} \equiv (1/T) \sum_{t=T+1}^{2T} v(t)$
for another $T$ steps, where the spatial average
$v(t) \equiv (1/N) \sum_{n=1}^N  v_n^t$.
The traffic flux is then computed by $\Phi \equiv \rho v_{\rm avg}$.
We confirm that $T=$ 30 000 is long enough to ensure that the system approaches
the steady state.
The slow increase and the decrease of $\rho$ are
implemented as follows: in order to disturb the flow as little as possible,
the addition of a single vehicle is done at the position where
the distance between two cars ($x_{n+1} - x_n$) is maximum.
When a car is removed from the road, it is done for the slowest car,
in order to make the jam dissolve more easily.

Figure~\ref{fig:rhoPhi} exhibits the (global) fundamental diagram, the flux $\Phi$
versus the vehicular density $\rho$. The upper branch of the curve
is obtained when $\rho$ is slowly increased. When $\rho = 50$(veh/km)
is approached from below, we begin to decrease $\rho$ as described
above.  The dotted blue curve marked as "Mechanism On" in Fig.~\ref{fig:rhoPhi}
will be explained below.
As is well-known, the fundamental diagram exhibits hysteresis
behavior: $\Phi$ follows different curves as $\rho$ is increased and
decreased. The existence of the hysteresis loop indicates that once the
traffic jam is formed, it is quite difficult to dissolve the jam
unless the vehicular density is significantly decreased. On the other hand, if we decrease
$\rho$ when the system is at point {\bf c} in Fig.~\ref{fig:rhoPhi}, we confirm
that the system can easily go back to the F phase. In other words, in the
present model, there is no hysteresis behavior between the F phase and the S
phase, which in turn indicates that the transition between S and F is
not discontinuous in our study. 
In comparison, some of existing CA model studies have produced 
a continuous F-S transition~\cite{jiang,gao} as in the present work, 
while others supported the discontinuous transition~\cite{jiang2}.  

For better understanding of the characteristics of the three different phases,
the spatio-temporal traffic patterns are displayed in Fig.~\ref{fig:xt}:
each panel corresponds to six different points (a-f)
in the fundamental diagram (Fig.~\ref{fig:rhoPhi}). An advantage of
the traffic model in Ref.~\cite{hklee} is that we can easily
see the connection between driving behavior and the traffic flow.
In the F phase [{\bf a} in Fig.~\ref{fig:rhoPhi} and Fig.~\ref{fig:xt}(a)],
almost all drivers become optimistic and drive at the maximum speed allowed.
As $\rho$ is increased, the inhomogeneous S phase (b)
is clearly observed between (a) the F and (c) the homogeneous
S phases. A further increase of $\rho$ eventually results in (d) the
J phase. Interestingly, we observe that the downstream front of the
jam and the downstream front of the synchronized region move at different
speeds, as can be compared in Figs.~\ref{fig:xt}(b) and 
(d)~\cite{footnote2}.
We emphasize that
the traffic flow patterns in Fig.~\ref{fig:xt} are obtained after long periods:
under the periodic boundary condition adopted in the present study,
once the jamming occurs, it is not spontaneously dissolved either to the F
phase [from (f) to (a)] or to the S phase [from (e) to (c)].
The measured speed ($\approx$ 50 km/h) of the downstream front of the 
synchronized region in Fig.~\ref{fig:xt}(b) is much higher than
the corresponding speed ($\approx$ 18.4 km/h) for the jammed region 
in Fig.~\ref{fig:xt}(e) and (f). In comparison, the improved Kerner's
three-phase model in Ref.~\cite{jiang3} has produced a realistic speed 
(5 cells per time unit corresponding to 8.5km/h) of the downstream front of the synchronized flow, 
while in Fig.~3 of Ref.\cite{jin} the corresponding speed is measured to be 
roughly about 32km/h.
We remark that the somehow unrealistically high speed of the downstream front
of the synchronized flow in Fig.~\ref{fig:xt}(b) could be an artifact 
of the present model. Likewise, the absence of the hysteresis behavior
around S-F transition could also be model dependent.

\begin{figure*}
\includegraphics[width=0.9\textwidth]{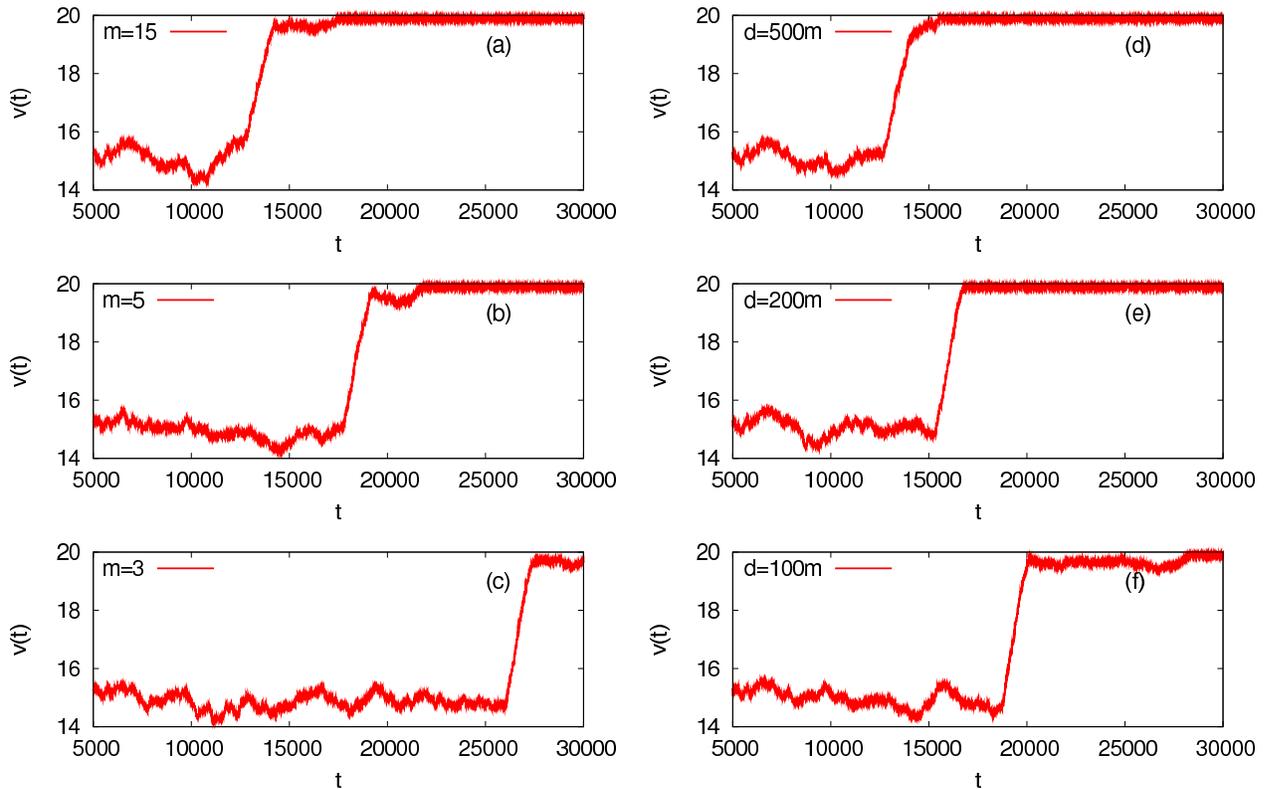}
\caption{(Color online) The average speed $v(t)$ versus the time $t$. We start from
the J phase (at {\bf f} in Figs. 1 and 2), and turn on our jam dissolving mechanism
(see text) at $t=5000$. (a)-(c) The first method based on the speed of the
$m$-th leading vehicle. (d)-(f) The second method based on the speed of the vehicle
at the geographic distance $d$.}
\label{fig:fig3}
\end{figure*}

The main objective of this study is to suggest a simple way to dissolve the
traffic jam within the limitation of the present CA model. 
If strong enforcement by the central authority is allowed,
the most efficient way to have maximum traffic flow is to
directly control the vehicular density $\rho$ so that traffic flow is tuned
toward the point {\bf a} in Fig.~\ref{fig:rhoPhi}. In this iron-fist solution, more cars
are allowed to enter the road if $\rho < \rho_c \approx 20$(veh/km), while
all the entrances are blocked if $\rho > \rho_c$. Although not realistic, this centralized
way of maximizing traffic flow could be a very efficient way thanks to
the absence of the hysteresis behavior between the F and the S phases.
In other words, although the central authority fails to detect the maximum-flow
condition $\rho = \rho_c$ and thus the synchronized flow already starts,
a simple reduction of $\rho$ will put the traffic back to the free-flow state.
In the real-world traffic, however, it is not plausible that drivers are
willing to accept the centralized control. We believe that the above
consideration raises an interesting issue in the game theoretic situation: near
$\rho = \rho_c$, the flow is maximum and every driver will complain much if she
is not allowed to enter the road which her leading vehicle just entered
in a few seconds earlier. For her own sake, she is better off entering the road
against the will of the central authority. However, in the public perspective, the
overall traffic flow becomes worse by allowing her to get in. A similar situation
has been termed as the price of anarchy in the study of urban car traffic~\cite{poa}.

It is to be noted that in real traffic situations, feedback control of traffic
flux at ramps is better suited to dissolve the traffic jam than the control of the
global average vehicular density. The spatial inhomogeneity and the unavoidable
time lag between the start of the control and the system's response require
adaptive control on ramps. In this regard, "ramp-metering" has already
been well studied and applied in real highway systems~\cite{ramp}.

We suggest below a noncentralized way to dissolve the traffic jam which is
based on local traffic information. Traffic flow can be of course improved by
building a new road. However, the construction of new roads is not
a practical alternative anymore in most countries~\cite{acc}.
Accordingly, a recent focus towards better traffic flow has
been made regarding how we can use existing roads in a more intelligent
way. One of the promising directions for achieving better
traffic flow is by using the so-called adaptive cruise control~\cite{acc,HK-acc2}.  The
basic idea behind this is that each car can be equipped with a detector which
senses local traffic information and sends it to nearby cars.  This
gathered information can then be used to decide a driving strategy.
In the same spirit, we propose below a jam dissolving mechanism which 
is based on downstream traffic information.

The robustness of the jam phase indicates that the incoming flux to the jam is
not smaller than the outgoing flux.
Accordingly, if an appropriate softening of the myopic behavior of
jam-approaching drivers is made, it could reduce the incoming flux,
resulting in the eventual dissolution of the jam.
From this reasoning, we propose a simple way to dissolve the
jam: if a vehicle is approaching the upstream jam front, the driving behavior
becomes defensive.
In this regard,
the use of the model in Ref.~\cite{hklee} has a great benefit since the model
already contains the dynamic variable $\gamma$ which describes the driving
behaviors. Accordingly, our method is written as
\begin{itemize}
\item If the $m$th leading vehicle has speed lower than the threshold value $v_{th}$, (i.e.,
      if $v_{n+m}^t <  v_{th}$), the driver becomes defensive ($\gamma_n^t = 1$).
\end{itemize}
In the broad range of $m$ and $v_{th}$ we indeed observe that the traffic jam
is dissolved.  We then try to optimize the values of $m$ and $v_{th}$ to make
the dissolution of the jam happen as quickly as possible. As a result, we find
that when $v_{th}= 8 (\approx 43$ km/h) and $10 \leq m \leq 30$ the jam is
dissolved faster than the use of other values of $v_{th}$ and $m$. 
In Fig.~\ref{fig:fig3}(a)-(c), we display the jam dissolving behavior for 
various values of $m$: even at $m=3$ the jam is found to dissolve
eventually, as shown in Fig.~\ref{fig:fig3}(c).  It is noteworthy that the congested traffic state is characterized
by $v < $40 km/h in Ref.~\cite{acc}, similar to our optimized value of
$v_{th}$.  We also use a modified version of our method in which the geographic
distance $d$ to the downstream preceding vehicle is used instead of $m$: If the
farthest vehicle in the downstream direction at a distance shorter than $d$
has a speed lower than $v_{th}$, $\gamma$ is set to one (defensive driving).
We find that the jam is dissolved in a broad range of $d$ for $d \gtrsim 50$
corresponding to $d \gtrsim 75$ m [see Fig.~\ref{fig:fig3}(d)-(f)]. 
We also observe that if the fraction of
vehicles with our method turned on is larger than about 80\%, the jam is eventually dissolved
for both types of our methods.

\begin{figure*}
\includegraphics[width=0.9\textwidth]{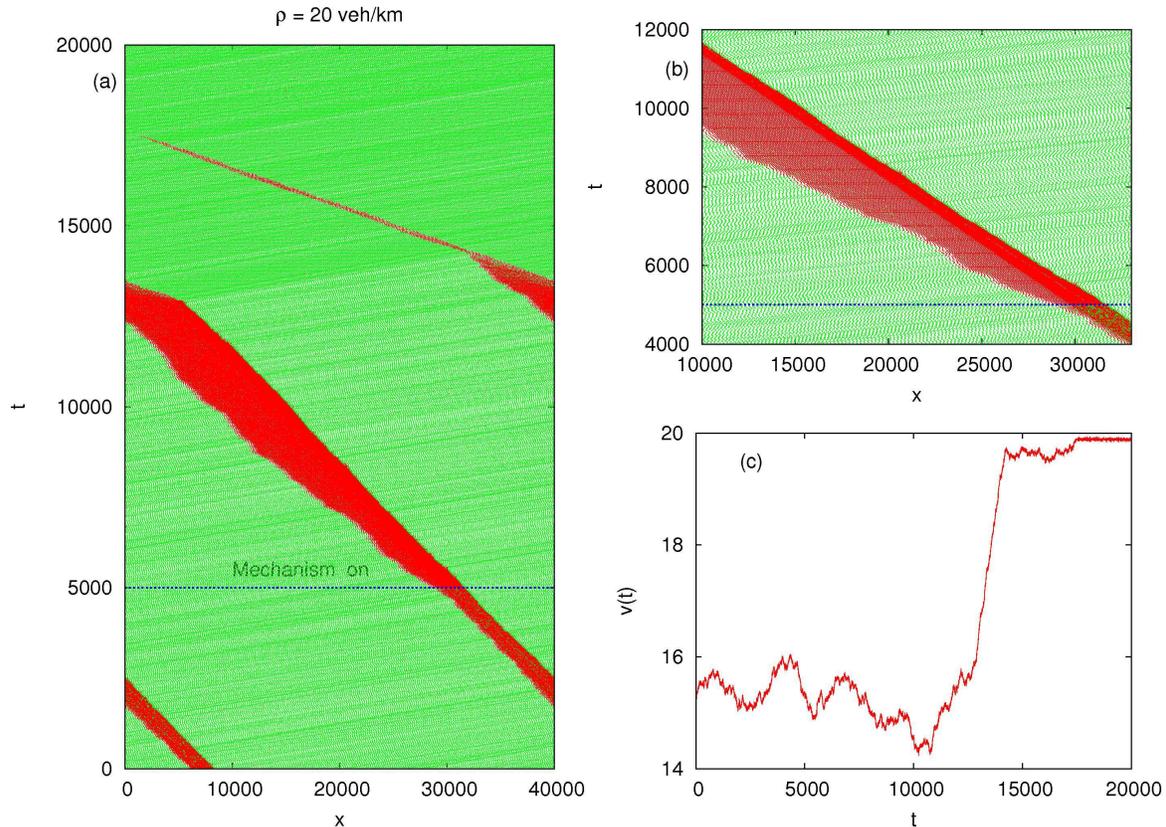}
\caption{(Color online) (a) Spatio-temporal ($x$,$t$) pattern of the traffic
flow starting from the J phase at point {\bf f} in Figs.~\ref{fig:rhoPhi} and
\ref{fig:xt}.  After the jam dissolving mechanism proposed in this study is
turned on at $t=$5 000, more drivers become defensive (red dots) in the
intermediate times, and eventually the traffic jam is dissolved for all drivers
become optimistic (green dots). (b) The expansion of a part of (a) to show more
clearly that the jammed region becomes narrower after $t=$ 5000.  (c) The
spatial average speed $v(t)$ versus time $t$.}
\label{fig:fig4}
\end{figure*}

\begin{figure*}
\includegraphics[width=0.9\textwidth]{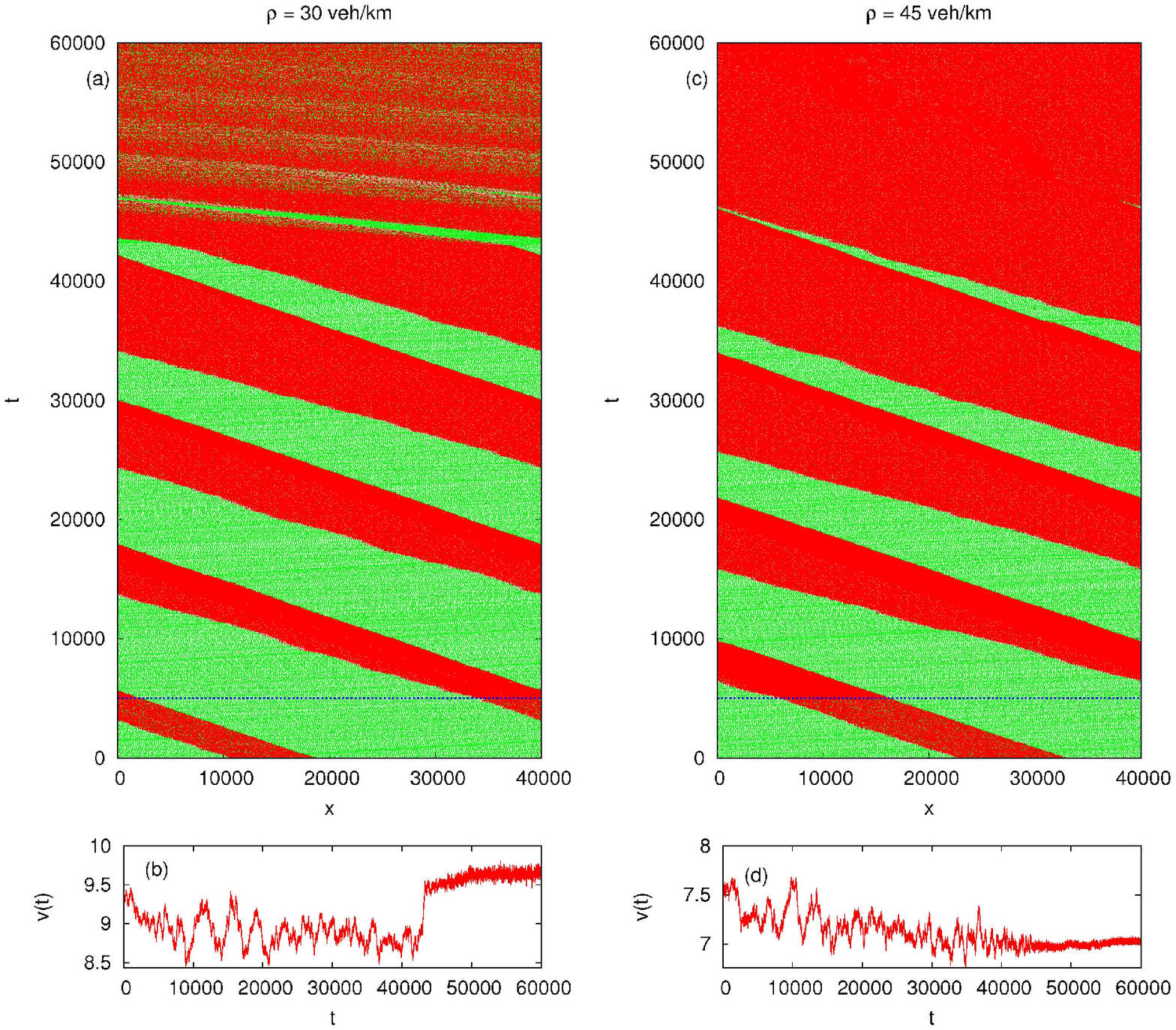}
\caption{(Color online) Spatio-temporal ($x$,$t$) pattern of the traffic flow 
starting from the J phase at (a) $\rho = 30$ and (c) $\rho = 45$. The 
average velocity $v(t)$ as a function of time $t$ for the corresponding
spatio-temporal pattern is also shown in (b) and (d). It is to be noted
that at $\rho = 30$, our jam dissolving mechanism improves the flow, while
at $\rho = 45$ it makes the flow worse (see Fig.~\ref{fig:rhoPhi}).}
\label{fig:fig5}
\end{figure*}

Figure~\ref{fig:fig4} exhibits the dissolution of the jam in the
spatio-temporal plot ($x$-$t$) [(a) and (b)], and (c) the spatial average
$v(t)$ as a function of time $t$, at $\rho = 20.0$(veh/km).  The initial state
at $t=0$ is obtained from the long-time simulation at point {\bf f} in
Figs.~\ref{fig:rhoPhi} and \ref{fig:xt}, and the mechanism explained above 
is on at $t = $5 000 with the parameters $m=15$ and $v_{th} = 8$.  It is
clearly seen that as time goes on, more and more jam-approaching drivers become
defensive [Fig.~\ref{fig:fig4}(a)], and the jammed region with higher local
density becomes narrower [Fig.~\ref{fig:fig4}(b)], as the upstream jam front
eventually catches up the downstream jam front at $t \sim 14 000$ 
[Fig.~\ref{fig:fig4}(a)].
Interestingly, a narrow track of defensive drivers persists for a quite long time,
until all drivers become optimistic at around $t=$ 18 000.  In
Fig.~\ref{fig:rhoPhi}, we also include the curve when our method of using
additional downstream traffic information is turned on at all times. The key
observations in Fig.~\ref{fig:rhoPhi} are as follows: (i) The hysteresis
behavior disappears and thus increase and decrease of $\rho$ result in the same
curve in the fundamental diagram.  (ii) The use of our method greatly improves
the traffic flow in a broad range of vehicular densities for $\rho \lesssim 26$,
and effectively removes the jam phase in the lower branch. (iii) When $26
\lesssim \rho \lesssim 31$, our method still helps but the traffic flow is
worse than the upper branch without it. (iv) Interestingly, when $\rho \gtrsim
31 $, the traffic flow with our method on is {\it worse} than the jam phase
without it.  (v) However, the jam that is robust in the broad range of $20 \leq \rho
\leq 30$ is efficiently dissolved by applying our simple method, which leads to
a considerable flux increase compared to that of the jam phase in this
density region.

For comparisons, we also display in Fig.~\ref{fig:fig5} the spatio-temporal
patterns and the average velocities versus time  at $\rho = 30$ [(a) and (b)],
and $\rho = 45$ [(c) and (d)], respectively.  The mechanism is turned on 
at $t=$ 5 000 as in Fig.~\ref{fig:fig4}.  At
$\rho = 30$, the congested region eventually spreads out and the average
velocity in the steady state becomes larger than the value at the lower branch
in Fig.~\ref{fig:rhoPhi} [see Fig.~\ref{fig:fig5}(b)]. In contrast, when the
vehicular density is too high as in Fig.~\ref{fig:fig5}(c), the steady state
value of the average velocity is smaller than the one with the mechanism off
[see Fig.~\ref{fig:fig5}(d) and (iv) above].

In summary, we have used the single-lane traffic model proposed in
Ref.~\cite{hklee} to study the nature of three traffic phases. As the vehicular
density is slowly increased and decreased, hysteresis behavior is observed
in the fundamental diagram for the flux and the density.  Within the 
limitations
of the used CA model, we propose a simple method to dissolve the jam, in which
driving behavior is forced to become defensive if the leading vehicle a few
cars ahead is dangerously slow.  We believe that the main conclusion of jam
dissolution via slowing down the jam-approaching vehicles by informing them
of the speeds of leading vehicles could be general. However, the validity of our
results needs to be checked more carefully in further studies.
Our proposed mechanism dissolves the jam since it reduces the inflow 
without altering the outflow from the jam. Combined with the
complementary method of increasing the outflow by using 
automated cruise control, the elimination of the jam can be made
more effective.

B.J.K. was supported by Basic Science Research Program through the 
National Research Foundation of Korea (NRF)
funded by the Ministry of Education, Science and Technology (2010-0008758),
and H.K.L by grant (07-innovations in techniques A01) from the National
Transportation Core Technology Program funded by Ministry of Land, Transport and
Maritime Affairs of Korean government.


\begin{thebibliography}{1}
\bibitem{footnote2} When the ringroad system (or the road with the periodic
boundary condition) is in the steady state, the upstream and the
downstream fronts move at the same speed. In reality, the speed of
the upstream front depends on the local influx of vehicles at the front
and thus the speed of the downstream front is more fundamental quantity.
\end{thebibliography}
\end{document}